\documentclass[twocolumn,superscriptaddress,floatfix,preprintnumbers, nofootinbib,hyperref]{revtex4-2} 
\pdfoutput=1
\usepackage[colorlinks=true,breaklinks=true]{hyperref}
\usepackage[normalem]{ulem}
\usepackage[utf8]{inputenc}
\hypersetup{allcolors=[rgb]{0.0 0.0 0.6},linkcolor=[rgb]{0.75 0.05 0.05}}
\usepackage{amsmath,amssymb}
\usepackage{epsfig}  
\usepackage{graphicx}   
\usepackage{slashed}       
\usepackage{tikz}
\usepackage{tikz-feynman}
\usepackage{url}
\usepackage{color}
\usepackage{multirow}
\usepackage{comment}
\usepackage{amssymb}

\DeclareMathOperator{\cm}{cm}

\DeclareMathOperator{\eV}{eV}

\newcommand{\beq}{\begin{equation}}
\newcommand{\eeq}{\end{equation}}

\allowdisplaybreaks

\bibpunct{[}{]}{,}{n}{}{,}

\begin{document}

\title{Glueball Dark Matter revisited}

\author{Pierluca Carenza}
\affiliation{The Oskar Klein Centre, Department of Physics, Stockholm University, Stockholm 106 91, Sweden
}

\author{Roman Pasechnik}
\affiliation{
 Department of Astronomy and Theoretical Physics, Lund University\\
 SE-223 62 Lund, Sweden
}

\author{Gustavo Salinas}
\affiliation{The Oskar Klein Centre, Department of Physics, 
Stockholm University, Stockholm 106 91, Sweden
}

\author{Zhi-Wei Wang}\email{Corresponding author.\\
zhiwei.wang@thep.lu.se}
\affiliation{School of Physics, The University of Electronic Science and Technology of China,\\
 88 Tian-run Road, Chengdu, China}

\smallskip

\begin{abstract}
We revisit the possibility that Dark Matter is composed of stable scalar glueballs of a confining dark ${\rm SU}(3)$ gauge theory coupled only to gravity. The relic abundance of dark glueballs is studied for the first time in a thermal effective theory accounting for strong-coupling dynamics. An important ingredient of our analysis is the use of an effective potential for glueballs that is fitted by lattice simulations. We predict the relic abundance to be in the range  $0.12\zeta_{T}^{-3}\Lambda/(137.9\eV) \lesssim \Omega  h^{2}\lesssim 0.12\zeta_{T}^{-3}\Lambda/(82.7\eV)$, with $\Lambda$ being the confinement scale, $\zeta_{T}$ the visible-to-dark sector temperature ratio and the uncertainty is coming from the fit to lattice data. This prediction is an order of magnitude smaller than the existing glueball abundance results in the literature. Our framework can be easily generalised to different gauge groups and modified cosmological histories paving the way towards consistent exploration of strongly-coupled dark sectors and their cosmological implications.
\end{abstract}

\maketitle

\emph{Introduction---}Confining dark Yang-Mills sectors are often considered as a possible source of Cold Dark Matter (CDM) in the Universe. In the simplest case, such dark gauge sectors are decoupled from the Standard Model, except for the gravitational interaction. However, the strong self-interactions confine the gauge sector into composite objects such as glueballs, in the case of a dark sector only composed of dark gluons. In a minimal approach, the lightest composite state predicted by a pure strongly-coupled gauge theory, the scalar dark glueball, is extensively discussed in the literature as a possible natural CDM candidate~\cite{Carlson:1992fn,Faraggi:2000pv,Feng:2011ik,Boddy:2014yra,Soni:2016gzf,Kribs:2016cew,Acharya:2017szw,Dienes:2016vei,Soni:2016yes, Soni:2017nlm,Draper:2018tmh,Halverson:2018olu, forestell2017cosmological, forestell2017non-abelian} (see also Ref.~\cite{Dondi:2019olm} for a more general discussion including `dark hadrons' and Ref.~\cite{Das:2018ons} for phenomenology of generic late-time forming DM). As an important case of self-interacting DM, this type of DM enables a consistent description of the structure of the Universe at small scales, in particular, helping resolve the so-called missing satellite problem \cite{Mateo:1998wg} and the cusp-core problem in the CDM distribution at galactic scales \cite{Spergel:1999mh,de_Blok_2010}. Furthermore, strongly-coupled dark Yang-Mills theories resembling Quantum Chromodynamics (QCD) in the Standard Model are physically motivated (e.g., these sectors show up frequently in string compactifications~\cite{Gross:1984dd,Dixon:1985jw,Ibanez:1986tp,Cvetic:2004ui,Gmeiner:2005vz,Drummond:2008aq,Acharya:1998pm,Grassi:2014zxa,Georgi:2014sxa,Taylor:2015xtz,Halverson:2016vwx}) and a wealth of knowledge in non-perturbative QCD can be directly applied there. Note, since only a pure Yang-Mills theory has robust and clean results available from lattice simulations \cite{Panero:2009tv}, it has traditionally been the best starting point to study strongly coupled dark sectors rigorously.

With the presence of a first-order confinement-deconfinement phase transition at a critical temperature $T_c$ \cite{Panero:2009tv,Halverson:2020xpg,Huang:2020crf,Reichert:2021cvs,Kang:2021epo}, an analysis of relic abundance of this type of DM is nontrivial and requires a detailed knowledge of thermal field theory in a non-perturbative domain. The existing calculations predict that the relic abundance of dark glueballs overcloses the Universe for a confining sector with critical temperature above the eV-scale, if that sector is not significantly cooler than the SM thermal bath. When multiple dark gauge sectors are present, a situation ubiquitous in string theory, this becomes a serious problem for phenomenology~\cite{Halverson:2016nfq}. Therefore, a precise understanding of the cosmological generation of glueball DM, with the inclusion of strong-coupling effects, is necessary. In this work, we develop a novel approach to study the relic abundance of dark glueballs by using the well established low-energy effective model of glueball and gluon dynamics at finite temperatures \cite{Sannino:2002wb}. We further constrain the effective model parameters by means of lattice results such as thermodynamic quantities and observables of the gluon condensate at finite temperature. 

Our approach provides for the first time a rigorous theoretical treatment of the dark glueball dynamics yielding a prediction for the range of relic abundance  $0.12\zeta_{T}^{-3}\Lambda/(137.9\eV) \lesssim \Omega  h^{2}\lesssim 0.12\zeta_{T}^{-3}\Lambda/(82.7\eV)$, about an order of magnitude below the previous estimates in Refs.~\cite{Carlson:1992fn,Halverson:2016nfq}, depending on the visible-to-dark sector temperature ratio $\zeta_{T}$. We confirm the linear dependence of the relic abundance with the confinement scale which is the essence of dark glueball overproduction problem in the early Universe while the relic abundance itself is significantly reduced.

\emph{Glueball effective Lagrangian---}A first-principle's treatment of the $SU(N)$ confinement-deconfinement phase transition is a tough theoretical challenge which requires a consistent description of a deeply non-perturbative dynamics. Lattice simulations represent a valuable tool to study phase transitions in Yang-Mills theories with and without matter fields (e.g., see~Refs.~\cite{Boyd:1996bx,CP-PACS:1999eop,Panero:2009tv}). At the same time, other complementary approaches have been used to understand different aspects of the strong-coupling effects, such as effective models and the functional renormalization group~\cite{Meisinger:2001cq,Meisinger:2001fi,Dumitru:2000in,Agasian:1993fn,Campbell:1990ak,Simonov:1992bc,Sollfrank:1994du,Carter:1998ti,Schaefer:2001cn,Drago:2001gd,Renk:2002md,Pisarski:2001pe,KorthalsAltes:1999cp,Dumitru:2001xa,Wirstam:2001ka,Laine:1999hh,Sannino:2002re,Scavenius:2001pa,Scavenius:2002ru,DelDebbio:2002nb}. Here, we describe the dynamics of dark glueballs by means of an effective field theory~\cite{Sannino:2002wb}.

At non-vanishing temperatures $T$, the $\mathbb{Z}_N$ center of $SU(N)$ is a relevant global symmetry~\cite{Svetitsky:1982gs} and it is possible to construct a number of gauge invariant operators charged under $\mathbb{Z}_N$. The Polyakov loop is a remarkable example, defined as
\begin{equation}
 {\ell}\left(x\right)=\frac{1}{N}{\rm Tr}[{\bf L}]\equiv\frac{1}{N}{\rm Tr}
\left[{\cal
P}\exp\left[i\,g\int_{0}^{1/T}A_{0}(\tau, \mathbf{x})d\tau\right]\right]\,,   
\end{equation} 
where ${\cal P}$ denotes path ordering, $A_{0}$ is the time component of the vector potential associated with this gauge group, $g$ is the $SU(N)$ coupling constant and $(\tau, \mathbf{x})$ are Euclidean spacetime coordinates. The Polyakov loop is charged with respect to the center $\mathbb{Z}_N$ of the $SU(N)$ gauge group~\cite{Svetitsky:1982gs} under which it transforms as $\ell \rightarrow z \ell$ with $z\in \mathbb{Z}_N$. Since the expectation value of the Polyakov loop vanishes at temperatures below the critical one and it is non-zero at higher temperatures, it is typically used as an order parameter for the Yang-Mills confinement phase transition at temperature $T_c \sim \Lambda$~\cite{Svetitsky:1982gs}. This observation was exploited to model the phase transition in a mean field approach in terms of Polyakov loops known as the Polyakov Loop Model (PLM)~\cite{Pisarski:2001pe}. This model captures the essential features of confinement phase transition in $SU(N)$ theories with $N\geq2$ while PLM-inspired models were also proposed to understand physics of heavy-ion collisions at the RHIC collider~\cite{Scavenius:2001pa,Scavenius:2002ru}. In \cite{Huang:2020crf}, it has been shown that PLM can very well capture thermodynamic observables predicted by lattice simulations \cite{Panero:2009tv}.

At temperatures around $T_c$, one can treat the glueball field $\mathcal{H}$ and the Polyakov loop $\ell$ in a unified description, with an effective temperature-dependent potential given by~\cite{Sannino:2002wb}
\begin{equation}
V\left[\mathcal{H},\ell \right]
=\frac{\mathcal{H}}{2}\ln\left[\frac{\mathcal{H}}{\Lambda^4}\right]  + T^4 {\cal
V}\left[ \ell \right]+\mathcal{H}{\cal P}{\left[ \ell \right]} +V_{T}\left[\mathcal{H}\right] \,, 
\label{eq:potential}
\end{equation}
where the first term is the zero-temperature glueball potential which can be obtained via the constraint of trace anomaly \cite{Schechter:1980ak,Schechter:2001ts}, $\Lambda$ is the confinement scale of the theory, and ${\cal V}\left[ \ell \right]$ and ${\cal P}\left[ \ell \right]$ are assumed to be real polynomials in $\ell$ and invariant under $\mathbb{Z}_N$, with coefficients that depend on fits to lattice data. Thermal corrections are included in $V_{T}[\mathcal{H}]$, which might involve terms that are non-analytic in $\mathcal{H}$ \cite{Schaefer:2001cn}.

Note that (i) the potential in Eq. (\ref{eq:potential}) reduces to the glueball dynamics at low temperatures and follows the PLM in the hot phase, (ii) the glueball field $\mathcal{H}$ is a dimension four scalar field and (iii) the term that couples $\mathcal{H}$ and $\ell$ is the most general interaction term which can be constructed without spoiling the zero temperature trace anomaly (Eq.~(21) of Ref.~\cite{Schechter:2001ts}). 

In this simplified model we neglect the entire tower of heavier glueballs and pseudo-scalar glueballs and the infinite series of dimensionless gauge invariant operators with different charges under $Z_N$. Nevertheless this model describes the essential features of the Yang-Mills phase transition. Below the critical temperature $T_c$ the last term in Eq.~(\ref{eq:potential}) is negligible. Since the glueballs are relatively heavy compared to the $\Lambda$ scale their temperature contribution $V_T[\mathcal{H}]$ can also be disregarded in the first approximation \cite{Sannino:2002wb}. We leave a refined analysis accounting for thermal effects in the glueball potential for a future investigation.

In the opposite limit, $T\gg T_{c}$, in the deconfined phase, the term $T^4{\cal V}[\ell]$ dominates, i.~e. dark gluons are the dominant component. The precise relation between the confinement scale $\Lambda$ and the critical temperature of the phase transition $T_{c}$ depends mildly on the gauge group and matter structure of the theory and is determined by lattice simulations. In this paper, we consider $T_c \sim 1.61 \Lambda$ for SU(3) (see e.g.~Ref.~\cite{Lucini:2012wq} for arbitrary number of colors).
\begin{table}[t!]
\vspace{0.5cm}
	\begin{tabular}{|c|c|c|c|c|c|c|}
		\hline
		$a_0$ & $a_1$ & $a_2$ & $a_3$ & $a_4$ & $b_3$ & $b_4$ \\
		\hline
		 3.72 & $-$5.73 & 8.49 & $-$9.29 & 0.27 & 2.40 & 4.53 \\
		\hline
	\end{tabular}
	\caption{Parameters of the effective potential in Eq.~\eqref{eq:pote}.}
	\label{tab:best-fit}
\end{table}

We consider the following Lagrangian for the glueball and Polyakov loop degrees of freedom~\cite{Gomm:1985ut,Ouyed:2001fv,Sannino:2002wb}
\begin{equation}
    \mathcal{L}=\frac{c}{2}\frac{\partial_{\mu}\mathcal{H}\partial^{\mu}\mathcal{H}}{\mathcal{H}^{3/2}}-V[\mathcal{H},\ell]\,,
\end{equation}
where
\begin{equation}
    c = \frac{1}{2\sqrt{e}} \left(\frac{\Lambda}{m_{\rm gb}}\right)^2
\end{equation}
is a constant determined by the glueball mass $m_{\rm gb}$, that in the following is assumed to be $m_{\rm gb}=6\Lambda$ \cite{Curtin:2022tou}. The Polyakov loop is a non-dynamical order parameter and since it is assumed to be homogeneous in space, we ignore terms involving spatial derivatives of $\ell$. This corresponds to neglect the non-trivial dynamics of a first order phase transition, which proceeds via the formation of bubbles and their subsequent collisions. This could have a significant impact on the formation of glueballs, as observed in presence of matter (see e.g.~Refs.~\cite{Asadi:2021pwo,Asadi:2022vkc}). The kinetic term for the glueball field $\mathcal{H}$ is non-standard, as it can be inferred from its dimensionality. For this reason, we write the glueball field $\mathcal{H}$ in terms of a canonically normalised scalar field $\phi$ as $\mathcal{H}=2^{-8}c^{-2}\phi^{4}$, and from this point on we refer to $\phi$ as the glueball field. It evolves according to the effective Lagrangian
\begin{equation}
\begin{split}
     \mathcal{L}&=\frac{1}{2}\partial_{\mu}\phi\partial^{\mu}\phi-V[\phi,\ell]\,, \\
     V[\phi,\ell]&=\frac{\phi^{4}}{2^{8}c^2}\left[2\ln\left(\frac{\phi}{\Lambda}\right)-4\ln2-\ln c\right]+\\
    &\quad+\frac{\phi^{4}}{2^{8}c^2}\mathcal{P}[\ell]+T^4 {\cal V}\left[ \ell \right]\,,\\
    \mathcal{P}[\ell]&=c_1|\ell|^{2}\,,\\
      \mathcal{V}[\ell]&=-\frac{b_{2}(T)}{2}|\ell|^{2}+b_{4}|\ell|^{4}-b_{3}(\ell^{3}+(\ell^{*})^{3})\,,\\
    b_{2}(T)&=\sum_{i=0}^{4}a_{i}\left(\frac{T_{c}}{T}\right)^{i}\,,
\end{split}
\label{eq:pote}
\end{equation}
where we have kept only the lowest order in $\mathcal{P}[\ell]$ satisfying the symmetries. The Polyakov loop potential $\mathcal{V}[\ell]$ is determined from symmetry arguments combined with fits to lattice thermodynamic quantities. Our choice here is taken from Ref.~\cite{Huang:2020crf} and the numerical values of the constants are reported in Tab.~\ref{tab:best-fit}, for clarity. 
\begin{figure}[t!]
	\centering
	\includegraphics[width=0.95\columnwidth]{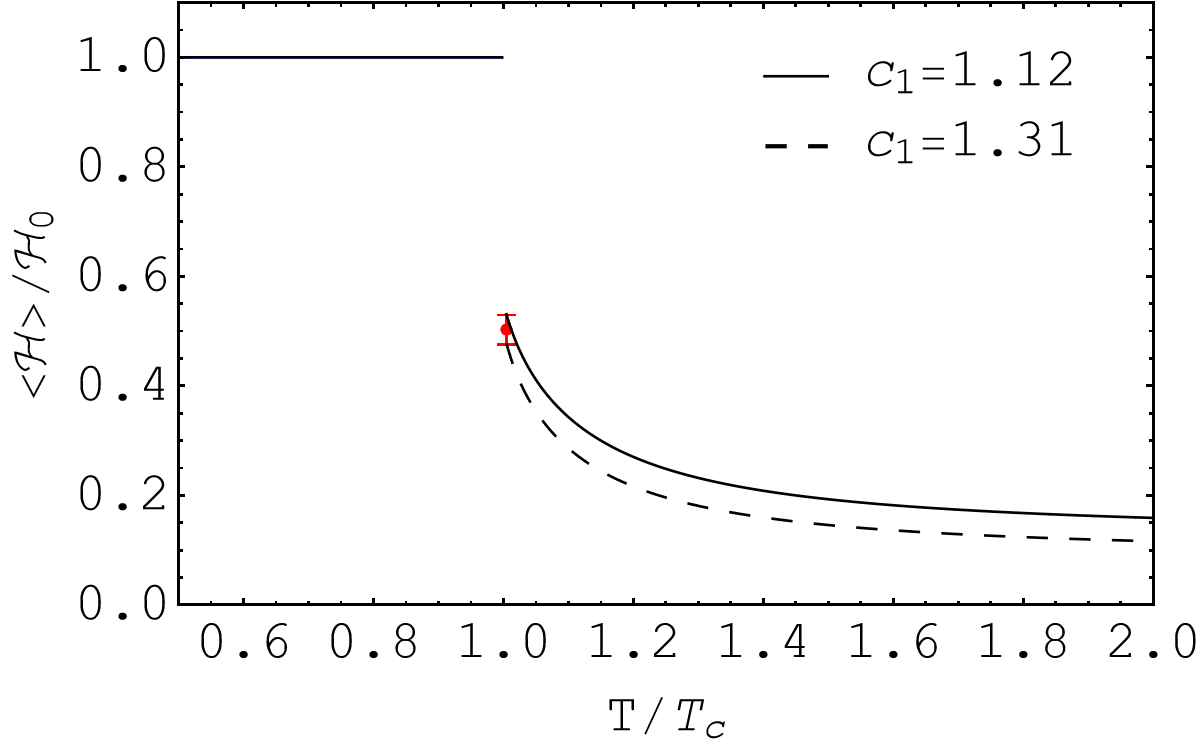}   	
	\caption{Vacuum expectation value of the glueball field $\mathcal{H}$ as a function of temperature. The field is normalized to its value in the confined phase.	The discontinuity at $T=T_{c}$ is characteristic of a first-order phase transition and the value of the jump depends on the parameter $c_{1}$, whose limiting values shown in this plot are obtained by a comparison with the lattice data~\cite{DElia:2002hkf}. The values shown correspond to $1\sigma$ uncertainty range. We do not use the lattice data for higher temperatures in the comparison, as our model neglects thermal corrections, which are increasingly relevant above $T_c$.}
	\label{fig:Vphi}
\end{figure} 

\emph{Temperature dependence of the Polyakov loop---}The Lagrangian in Eq.~\eqref{eq:pote} describes the evolution of the glueball-dark gluon system across the phase transition. This effective description is expected to be valid in a broad temperature range, except when the temperature is large $T\gg T_c$, where $V_T[\mathcal{H}]$ needs to be included. Since the Polyakov loop is a non-dynamical degree of freedom, its temperature evolution is determined by the location of the minimum in the effective potential. Being the order parameter of the phase transition, $\ell$ approaches $1$ at high temperatures and vanishes for temperatures below the critical one. The stationary points of $\ell$ are $\ell=0$ and
\begin{equation}
    |\ell_{\pm}|=\cfrac{3b_{3}}{4b_{4}}\left(1\pm\sqrt{1+\cfrac{512\,b_{2}(T)b_{4}- 4\cfrac{c_1b_{4}}{c^2}\left(\cfrac{\phi}{T}\right)^{4}}{1152\,b_{3}^{2}}}\right)\,,
\end{equation}
representing two minima, $\ell=0$ and $\ell=\ell_{+}$, separated by a maximum in $\ell=\ell_{-}$. The solution $\ell=0$ denotes the confined phase and it is a global minimum only for temperatures below the critical temperature. In the deconfined phase, the solution $\ell=0$ becomes metastable and $\ell=\ell_{+}$ becomes the global minimum. The Polyakov loop is then ``integrated out'' using its equation of motion $\ell=\ell(\phi,T)$, giving rise to a potential for the glueball field in the form $V[\phi,T]=V[\phi,\ell(\phi,T)]$. Moreover, we set the zero-point energy of the glueball field to zero in order to properly describe glueballs as matter. The evolution of the glueball minimum in this new potential is shown in Fig.~\ref{fig:Vphi} in terms of the field $\mathcal{H}$ and compared to lattice simulations. Below $T_c$, $\langle\mathcal{H}\rangle$ is constant with temperature and it discontinuously jumps to a lower value right above the critical temperature. We match the size of the discontinuity predicted in our potential to a result from lattice, given in Ref.~\cite{DElia:2002hkf} (the red point in Fig.~\ref{fig:Vphi}). This constraint is enough to impose limitations on the value of $c_{1}$ in Eq.~\eqref{eq:pote}, the glueball-Polyakov loop coupling. We found this value to be $c_{1}=1.225\pm0.19$ at 95$\%$ CL. The associated uncertainty of $\sim20\%$ dominates the uncertainty in the glueball relic abundance in our analysis, such that $\sim\mathcal{O}(3\%)$ uncertainties on the fitting parameters in Tab.~\ref{tab:best-fit} have been ignored.

\emph{Cosmological evolution of the glueball field---}Thanks to the previous discussion, we are left with a relatively simple recipe to describe the glueball field dynamics across the phase transition. Note that the evolution can be treated as completely classical, since the effective Lagrangian in Eq.~(\ref{eq:pote}) fully accounts for quantum effects at tree level.

In a first approximation, the glueball field is homogeneous and evolves in an expanding Friedmann-Lemaitre-Robertson-Walker (FLRW) Universe. The Klein-Gordon equation for a field in a FLRW metric reads
\begin{equation}
    \ddot{\phi}+3H\dot{\phi}+\partial_{\phi}V[\phi,T]=0\,,
    \label{eq:KG}
\end{equation}
where the Hubble parameter $H$ when glueballs form is approximately determined by the SM content of the Universe, as it is assumed to have more degrees of freedom than the confining dark sector and, if there are no interactions with the SM, this sector is colder than the SM thermal bath. We denote the visible-to-dark sector temperature ratio by $\xi_{T}$. The photon temperature $T_{\gamma}$ determines the Hubble parameter $H$ and can be taken as a time variable in Eq.~\eqref{eq:KG} by using 
\begin{equation}
    t=\frac{1}{2}\sqrt{\frac{45}{4\pi^{3}g_{*,\rho}(T_{\gamma})}}\frac{m_{P}}{T_\gamma^{2}}\,,
\end{equation}
where $m_{P}$ is the Planck mass and $g_{*,\rho}(T_\gamma)$ is the number of degrees of freedom of the SM bath at temperature $T_\gamma = \xi_T T$. Note that the dark sector temperature $T$ is the one that governs the phase transition, i.e.~entering in Eq.~\eqref{eq:potential}. In terms of this variable Eq.~\eqref{eq:KG} reads
\begin{equation}
      \frac{4\pi^{3}g_{*,\rho}}{45m_{P}^{2}}\xi_T^4 T^{6}\frac{d^{2}\phi}{dT^{2}}+\frac{2\pi^{3}}{45m_{P}^{2}}\frac{dg_{*,\rho}}{dT}\xi_T^4T^{6}\frac{d\phi}{dT}+\partial_{\phi}V[\phi,T]=0\,,
    \label{eq:KG2}
\end{equation}
where the second term can be neglected for a large range of temperatures as $g_{*,\rho}$ is constant except at a few isolated events (the QCD phase transition, for example). We consider it as a free parameter and take $g_{*,\rho}=100$, which has very little impact on our final result. The visible-to-dark sector temperature ratio can be absorbed in an effective Planck mass, $M \equiv m_P/\xi_T^2$.

The non-perturbative dynamics of the system is encoded in Eq.~\eqref{eq:KG2} and, after the phase transition, we assume that the energy density stored in the glueball field gives precisely the DM relic density. From the particle physics point of view the evolution can be described as follows. In the deconfined phase the Universe is populated by dark gluons that form glueballs at the phase transition, thanks to the interaction term in Eq.~\eqref{eq:pote}. When the phase transition is completed, DM glueballs populate the Universe and interact with each other following the potential in Eq.~\eqref{eq:pote}, corresponding to interactions in the form $(\phi-\phi_{\rm min})^{n}$ for $n=2, 3...$, with $\phi_{\rm min}$ being the value of the field at the minimum of the potential. The importance of the higher-$n$ terms depends on the displacement of $\phi$ from its minimum, which is a measure of the glueball density. If, for example, $\phi$ is very close to its minimum, only the quadratic term is relevant, which is equivalent to having a massive free field. On the other hand, large amplitudes (i.e.~larger densities) for $\phi$ require increasingly more non-linear interaction terms (see also Refs.~\cite{Yamanaka:2019aeq,Yamanaka:2019yek}).

\begin{figure}[t!]
	\centering
	\includegraphics[width=0.95\columnwidth]{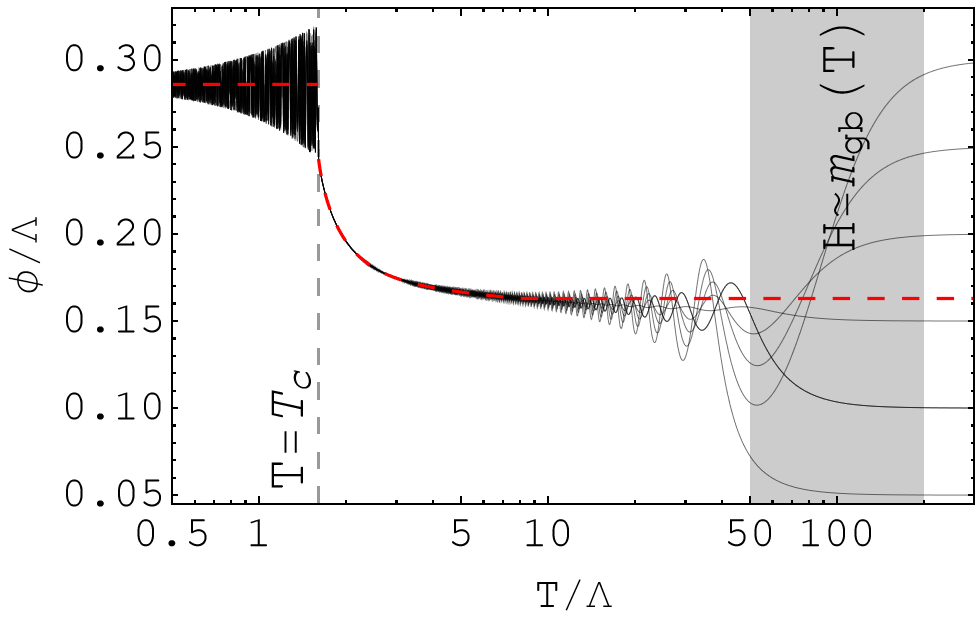}   	
	\caption{Evolution of the glueball field for a phase transition scale $\Lambda=10^{-5}m_{P}$, $c_{1}=1.225$ and different initial conditions. The grey region indicates the phase in which the glueball mass is comparable with the Hubble parameter, $H\simeq m_{\rm gb}(T)$.	The vertical dashed line marks the phase transition at $T_{c}=1.61\Lambda$. The red dashed line shows the evolution of the minimum of the glueball potential.}
	\label{fig:phit}
\end{figure}

In Fig.~\ref{fig:phit} we show the evolution of the glueball field as a function of temperature, starting from different initial conditions set in the deconfined phase. In the very early stage, the field evolution is dominated by the Hubble friction and it remains frozen until $H$ becomes comparable to the temperature-dependent effective glueball mass in the deconfined phase, represented by the gray region labelled as $H\simeq m_{\rm gb}(T)$. This happens at a temperature $T_{\rm osc} \sim \sqrt{M \Lambda}$, when the field starts to oscillate around the minimum of the potential, shown as a dashed red line in Fig.~\ref{fig:phit}, with a damped amplitude. We take $T_{\rm osc} \gg T_c$, as $M \gg \Lambda$, unless the confinement scale is close to the Planck scale or the dark sector is very cold. Therefore, the oscillations of the glueball field in the deconfined phase have enough time to decay, regardless of initial condition, and $\phi$ just follows the minimum of the potential (with damped oscillations of small amplitude but with an increasing average speed) until the phase transition occurs at $T_c$ (see Fig.~\ref{fig:phit}). At the critical temperature, the value of the Polyakov loop jumps discontinuously, causing a discontinuous jump in the minimum of the glueball potential, as shown in Fig.~\ref{fig:Vphi}, generating oscillations with a high initial velocity that wash out any dependence on initial conditions at $T>T_{c}$.

\emph{Glueball relic density---}In the confined phase, $\phi$ is displaced enough from its minimum to allow for annihilation of $n$ glueballs into $m<n$ glueballs, $n\to m$, which is possible because of the $n+m$-th order interaction term in the Lagrangian. As the glueball number density decreases, all the higher order $n\to m$ processes become less efficient until the only efficient number-changing process is $3\to 2$. Note that the $3\to1$ and $2\to 1$ processes are prohibited due to kinematic constraints arising from the energy conservation. The $3\to 2$ interactions are precisely the ones determining the relic abundance of glueballs when $\Gamma_{3\to2}<H$. The evolution is that of a simple damped oscillator in a non-linear potential, and the energy stored in these oscillations around $\phi_{\rm min} \approx 0.28\Lambda$ corresponds to the relic DM abundance, namely, $\Omega h^{2}=\rho/\rho_{c}$, where the critical density is $\rho_{c}=1.05\times10^{4}\eV\cm^{-3}$, and
\begin{equation}
    \rho=\frac{2\pi^{3}}{45}g_{*,\rho}(T)\frac{T^{6}}{M^{2}}\left(\frac{d\phi}{dT}\right)^{2}+V[\phi]\,.
\end{equation}
This energy density scales as $\sim T^{3}$, as CDM, when the harmonic approximation is valid, i.e.~after the decoupling of $3\to2$ interactions. Numerically solving  Eq.~\eqref{eq:KG2} down to the temperature $T_{f}$, and below this temperature the evolution is simply determined by the cosmological expansion, for $\Lambda\lesssim 0.1\,M$, the energy density is given by $0.015 \lesssim T_{f}^{-3}\Lambda^{-1}\rho \lesssim0.020$ for $1.035<c_{1}<1.415$. In conclusion, the predicted glueball relic density is
\begin{equation}
0.12\zeta_{T}^{-3}\frac{\Lambda}{137.9\eV} \lesssim \Omega  h^{2}\lesssim 0.12\zeta_{T}^{-3}\frac{\Lambda}{82.7\eV} \,,
\end{equation}
and this result should be compared to $\Omega  h^{2}\sim0.12 \,\zeta_{T}^{-3}\Lambda/5.45\eV$~\cite{Carlson:1992fn}, which overestimates the relic density by one order of magnitude. This difference is due to two main concurrent effects. The first one is an overestimation of the energy stored in the dark gluon field. In the literature, dark gluons are considered as radiation for all temperatures above the phase transition. In our approach, the energy density of dark gluons for temperatures right above the critical one strongly deviates (reduced by a factor $\sim 50$) from that of an ideal gas, in agreement with lattice results. The second effect is that glueballs do not redshift as CDM immediately after the phase transition, going through a phase in which their equation of state is $-1\lesssim p/\rho\lesssim0$, making them dilute slower than dust. The combination of these effects leads to the found discrepancy. We note also that thermal corrections increase the glueball relic density, by displacing the high-temperature minimum of a $\sim 10\%$ farther from the low-temperature minimum~\cite{Schaefer:2001cn}. We estimated an increase of the relic density up to $\sim80\%$ due to such thermal corrections, which will be subject of a future investigation.

A possible constraint on the model comes from the contribution of dark gluons to the effective number of relativistic species, constrained to be $\Delta N_{\rm eff}<0.35$ at the 95$\%$ CL~\cite{Planck:2018vyg}. A temperature ratio $\zeta_{T}\gtrsim 2$ is enough to evade this constraint. Therefore, a dark gauge sector interacting only via gravitational interactions with the SM and a confinement scale at the $\eV$ scale might explain the DM abundance without spoiling other cosmological observables. 

\emph{Discussion and conclusions---}In this work, we presented a new approach to calculate the glueball CDM relic density which includes the self-interactions in a non-perturbative fashion. We bridge the well-established thermal EFT with the existing lattice results to provide rigorous phenomenological predictions. Because of its generality, it is easy to apply this approach to different gauge groups, but in this work we considered only $SU(3)$ for the sake of clarity. Moreover, the method presented in this work is suitable for investigations of the glueball formation in modified cosmological histories, requiring only a simple modification of Eq.~\eqref{eq:KG2}, one of the main results of this work. Another interesting question is on the role of thermal effects in the glueball potential, that we neglected in this preliminary study. We postpone this study to a future work. Our work paves the road towards consistent exploration of strongly-coupled dark sectors and their cosmological implications.

\emph{Acknowledgements---}The work of P.C. and G.S. is supported by the European Research Council under Grant No.~742104 and by the Swedish Research Council (VR) under grants  2018-03641 and 2019-02337. R.P.~and Z.-W.~W.~are supported in part by the Swedish Research Council grant, contract number 2016-05996, as well as by the European Research Council (ERC) under the European Union's Horizon 2020 research and innovation programme (grant agreement No. 668679). Z.-W.~W.~is also supported by School of Physics, The University of Electronic Science and Technology of China. Z.-W.~W.~acknowledges Francecso Sannino for his nice comments.

\bibliographystyle{bibi}
\bibliography{biblio.bib}

\providecommand{\href}[2]{#2}\begingroup\raggedright\begin{thebibliography}{10}

\bibitem{Carlson:1992fn}
E.~D. Carlson, M.~E. Machacek and L.~J. Hall, \emph{{Self-interacting dark
  matter}}, \href{https://doi.org/10.1086/171833}{\emph{Astrophys. J.}
  {\bfseries 398} (1992) 43}.

\bibitem{Faraggi:2000pv}
A.~E. Faraggi and M.~Pospelov, \emph{{Selfinteracting dark matter from the
  hidden heterotic string sector}},
  \href{https://doi.org/10.1016/S0927-6505(01)00121-9}{\emph{Astropart. Phys.}
  {\bfseries 16} (2002) 451}
  [\href{https://arxiv.org/abs/hep-ph/0008223}{{\ttfamily hep-ph/0008223}}].

\bibitem{Feng:2011ik}
J.~L. Feng and Y.~Shadmi, \emph{{WIMPless Dark Matter from Non-Abelian Hidden
  Sectors with Anomaly-Mediated Supersymmetry Breaking}},
  \href{https://doi.org/10.1103/PhysRevD.83.095011}{\emph{Phys. Rev. D}
  {\bfseries 83} (2011) 095011}
  [\href{https://arxiv.org/abs/1102.0282}{{\ttfamily 1102.0282}}].

\bibitem{Boddy:2014yra}
K.~K. Boddy, J.~L. Feng, M.~Kaplinghat and T.~M.~P. Tait,
  \emph{{Self-Interacting Dark Matter from a Non-Abelian Hidden Sector}},
  \href{https://doi.org/10.1103/PhysRevD.89.115017}{\emph{Phys. Rev. D}
  {\bfseries 89} (2014) 115017}
  [\href{https://arxiv.org/abs/1402.3629}{{\ttfamily 1402.3629}}].

\bibitem{Soni:2016gzf}
A.~Soni and Y.~Zhang, \emph{{Hidden SU(N) Glueball Dark Matter}},
  \href{https://doi.org/10.1103/PhysRevD.93.115025}{\emph{Phys. Rev. D}
  {\bfseries 93} (2016) 115025}
  [\href{https://arxiv.org/abs/1602.00714}{{\ttfamily 1602.00714}}].

\bibitem{Kribs:2016cew}
G.~D. Kribs and E.~T. Neil, \emph{{Review of strongly-coupled composite dark
  matter models and lattice simulations}},
  \href{https://doi.org/10.1142/S0217751X16430041}{\emph{Int. J. Mod. Phys. A}
  {\bfseries 31} (2016) 1643004}
  [\href{https://arxiv.org/abs/1604.04627}{{\ttfamily 1604.04627}}].

\bibitem{Acharya:2017szw}
B.~S. Acharya, M.~Fairbairn and E.~Hardy, \emph{{Glueball dark matter in
  non-standard cosmologies}},
  \href{https://doi.org/10.1007/JHEP07(2017)100}{\emph{JHEP} {\bfseries 07}
  (2017) 100} [\href{https://arxiv.org/abs/1704.01804}{{\ttfamily
  1704.01804}}].

\bibitem{Dienes:2016vei}
K.~R. Dienes, F.~Huang, S.~Su and B.~Thomas, \emph{{Dynamical Dark Matter from
  Strongly-Coupled Dark Sectors}},
  \href{https://doi.org/10.1103/PhysRevD.95.043526}{\emph{Phys. Rev. D}
  {\bfseries 95} (2017) 043526}
  [\href{https://arxiv.org/abs/1610.04112}{{\ttfamily 1610.04112}}].

\bibitem{Soni:2016yes}
A.~Soni and Y.~Zhang, \emph{{Gravitational Waves From SU(N) Glueball Dark
  Matter}}, \href{https://doi.org/10.1016/j.physletb.2017.05.077}{\emph{Phys.
  Lett. B} {\bfseries 771} (2017) 379}
  [\href{https://arxiv.org/abs/1610.06931}{{\ttfamily 1610.06931}}].

\bibitem{Soni:2017nlm}
A.~Soni, H.~Xiao and Y.~Zhang, \emph{{Cosmic selection rule for the glueball
  dark matter relic density}},
  \href{https://doi.org/10.1103/PhysRevD.96.083514}{\emph{Phys. Rev. D}
  {\bfseries 96} (2017) 083514}
  [\href{https://arxiv.org/abs/1704.02347}{{\ttfamily 1704.02347}}].

\bibitem{Draper:2018tmh}
P.~Draper, J.~Kozaczuk and J.-H. Yu, \emph{{Theta in new QCD-like sectors}},
  \href{https://doi.org/10.1103/PhysRevD.98.015028}{\emph{Phys. Rev. D}
  {\bfseries 98} (2018) 015028}
  [\href{https://arxiv.org/abs/1803.00015}{{\ttfamily 1803.00015}}].

\bibitem{Halverson:2018olu}
J.~Halverson, B.~D. Nelson, F.~Ruehle and G.~Salinas, \emph{{Dark Glueballs and
  their Ultralight Axions}},
  \href{https://doi.org/10.1103/PhysRevD.98.043502}{\emph{Phys. Rev. D}
  {\bfseries 98} (2018) 043502}
  [\href{https://arxiv.org/abs/1805.06011}{{\ttfamily 1805.06011}}].

\bibitem{forestell2017cosmological}
L.~Forestell, D.~E. Morrissey and K.~Sigurdson, \emph{Cosmological bounds on
  non-abelian dark forces},
  \href{https://doi.org/10.1103/PhysRevD.97.075029}{\emph{arXiv: High Energy
  Physics - Phenomenology} (2017) }.

\bibitem{forestell2017non-abelian}
L.~Forestell, D.~E. Morrissey and K.~Sigurdson, \emph{Non-abelian dark forces
  and the relic densities of dark glueballs},
  \href{https://doi.org/10.1103/PhysRevD.95.015032}{\emph{Physical Review D}
  {\bfseries 95} (2017) 015032}.

\bibitem{Dondi:2019olm}
N.~A. Dondi, F.~Sannino and J.~Smirnov, \emph{{Thermal history of composite
  dark matter}}, \href{https://doi.org/10.1103/PhysRevD.101.103010}{\emph{Phys.
  Rev. D} {\bfseries 101} (2020) 103010}
  [\href{https://arxiv.org/abs/1905.08810}{{\ttfamily 1905.08810}}].

\bibitem{Das:2018ons}
A.~Das, B.~Dasgupta and R.~Khatri, \emph{{Ballistic Dark Matter oscillates
  above $\Lambda$CDM}},
  \href{https://doi.org/10.1088/1475-7516/2019/04/018}{\emph{JCAP} {\bfseries
  04} (2019) 018} [\href{https://arxiv.org/abs/1811.00028}{{\ttfamily
  1811.00028}}].

\bibitem{Mateo:1998wg}
M.~Mateo, \emph{{Dwarf galaxies of the Local Group}},
  \href{https://doi.org/10.1146/annurev.astro.36.1.435}{\emph{Ann. Rev. Astron.
  Astrophys.} {\bfseries 36} (1998) 435}
  [\href{https://arxiv.org/abs/astro-ph/9810070}{{\ttfamily
  astro-ph/9810070}}].

\bibitem{Spergel:1999mh}
D.~N. Spergel and P.~J. Steinhardt, \emph{{Observational evidence for
  selfinteracting cold dark matter}},
  \href{https://doi.org/10.1103/PhysRevLett.84.3760}{\emph{Phys. Rev. Lett.}
  {\bfseries 84} (2000) 3760}
  [\href{https://arxiv.org/abs/astro-ph/9909386}{{\ttfamily
  astro-ph/9909386}}].

\bibitem{de_Blok_2010}
W.~J.~G. de~Blok, \emph{The core-cusp problem},
  \href{https://doi.org/10.1155/2010/789293}{\emph{Advances in Astronomy}
  {\bfseries 2010} (2010) 1}.

\bibitem{Gross:1984dd}
D.~J. Gross, J.~A. Harvey, E.~J. Martinec and R.~Rohm, \emph{{The Heterotic
  String}}, \href{https://doi.org/10.1103/PhysRevLett.54.502}{\emph{Phys. Rev.
  Lett.} {\bfseries 54} (1985) 502}.

\bibitem{Dixon:1985jw}
L.~J. Dixon, J.~A. Harvey, C.~Vafa and E.~Witten, \emph{{Strings on
  Orbifolds}}, \href{https://doi.org/10.1016/0550-3213(85)90593-0}{\emph{Nucl.
  Phys. B} {\bfseries 261} (1985) 678}.

\bibitem{Ibanez:1986tp}
L.~E. Ibanez, H.~P. Nilles and F.~Quevedo, \emph{{Orbifolds and Wilson Lines}},
  \href{https://doi.org/10.1016/0370-2693(87)90066-9}{\emph{Phys. Lett. B}
  {\bfseries 187} (1987) 25}.

\bibitem{Cvetic:2004ui}
M.~Cvetic, T.~Li and T.~Liu, \emph{{Supersymmetric Pati-Salam models from
  intersecting D6-branes: A Road to the standard model}},
  \href{https://doi.org/10.1016/j.nuclphysb.2004.07.036}{\emph{Nucl. Phys. B}
  {\bfseries 698} (2004) 163}
  [\href{https://arxiv.org/abs/hep-th/0403061}{{\ttfamily hep-th/0403061}}].

\bibitem{Gmeiner:2005vz}
F.~Gmeiner, R.~Blumenhagen, G.~Honecker, D.~Lust and T.~Weigand, \emph{{One in
  a billion: MSSM-like D-brane statistics}},
  \href{https://doi.org/10.1088/1126-6708/2006/01/004}{\emph{JHEP} {\bfseries
  01} (2006) 004} [\href{https://arxiv.org/abs/hep-th/0510170}{{\ttfamily
  hep-th/0510170}}].

\bibitem{Drummond:2008aq}
J.~M. Drummond, J.~Henn, G.~P. Korchemsky and E.~Sokatchev, \emph{{Hexagon
  Wilson loop = six-gluon MHV amplitude}},
  \href{https://doi.org/10.1016/j.nuclphysb.2009.02.015}{\emph{Nucl. Phys. B}
  {\bfseries 815} (2009) 142}
  [\href{https://arxiv.org/abs/0803.1466}{{\ttfamily 0803.1466}}].

\bibitem{Acharya:1998pm}
B.~S. Acharya, \emph{{M theory, Joyce orbifolds and superYang-Mills}},
  \href{https://doi.org/10.4310/ATMP.1999.v3.n2.a3}{\emph{Adv. Theor. Math.
  Phys.} {\bfseries 3} (1999) 227}
  [\href{https://arxiv.org/abs/hep-th/9812205}{{\ttfamily hep-th/9812205}}].

\bibitem{Grassi:2014zxa}
A.~Grassi, J.~Halverson, J.~Shaneson and W.~Taylor, \emph{{Non-Higgsable QCD
  and the Standard Model Spectrum in F-theory}},
  \href{https://doi.org/10.1007/JHEP01(2015)086}{\emph{JHEP} {\bfseries 01}
  (2015) 086} [\href{https://arxiv.org/abs/1409.8295}{{\ttfamily 1409.8295}}].

\bibitem{Georgi:2014sxa}
H.~Georgi, G.~Kestin and A.~Sajjad, \emph{{Towards an Effective Field Theory on
  the Light-Shell}}, \href{https://doi.org/10.1007/JHEP03(2016)137}{\emph{JHEP}
  {\bfseries 03} (2016) 137} [\href{https://arxiv.org/abs/1401.7667}{{\ttfamily
  1401.7667}}].

\bibitem{Taylor:2015xtz}
W.~Taylor and Y.-N. Wang, \emph{{The F-theory geometry with most flux vacua}},
  \href{https://doi.org/10.1007/JHEP12(2015)164}{\emph{JHEP} {\bfseries 12}
  (2015) 164} [\href{https://arxiv.org/abs/1511.03209}{{\ttfamily
  1511.03209}}].

\bibitem{Halverson:2016vwx}
J.~Halverson, \emph{{Strong Coupling in F-theory and Geometrically
  Non-Higgsable Seven-branes}},
  \href{https://doi.org/10.1016/j.nuclphysb.2017.02.014}{\emph{Nucl. Phys. B}
  {\bfseries 919} (2017) 267}
  [\href{https://arxiv.org/abs/1603.01639}{{\ttfamily 1603.01639}}].

\bibitem{Panero:2009tv}
M.~Panero, \emph{{Thermodynamics of the QCD plasma and the large-N limit}},
  \href{https://doi.org/10.1103/PhysRevLett.103.232001}{\emph{Phys. Rev. Lett.}
  {\bfseries 103} (2009) 232001}
  [\href{https://arxiv.org/abs/0907.3719}{{\ttfamily 0907.3719}}].

\bibitem{Halverson:2020xpg}
J.~Halverson, C.~Long, A.~Maiti, B.~Nelson and G.~Salinas, \emph{{Gravitational
  waves from dark Yang-Mills sectors}},
  \href{https://doi.org/10.1007/JHEP05(2021)154}{\emph{JHEP} {\bfseries 05}
  (2021) 154} [\href{https://arxiv.org/abs/2012.04071}{{\ttfamily
  2012.04071}}].

\bibitem{Huang:2020crf}
W.-C. Huang, M.~Reichert, F.~Sannino and Z.-W. Wang, \emph{{Testing the dark
  SU(N) Yang-Mills theory confined landscape: From the lattice to gravitational
  waves}}, \href{https://doi.org/10.1103/PhysRevD.104.035005}{\emph{Phys. Rev.
  D} {\bfseries 104} (2021) 035005}
  [\href{https://arxiv.org/abs/2012.11614}{{\ttfamily 2012.11614}}].

\bibitem{Reichert:2021cvs}
M.~Reichert, F.~Sannino, Z.-W. Wang and C.~Zhang, \emph{{Dark confinement and
  chiral phase transitions: gravitational waves vs matter representations}},
  \href{https://doi.org/10.1007/JHEP01(2022)003}{\emph{JHEP} {\bfseries 01}
  (2022) 003} [\href{https://arxiv.org/abs/2109.11552}{{\ttfamily
  2109.11552}}].

\bibitem{Kang:2021epo}
Z.~Kang, J.~Zhu and S.~Matsuzaki, \emph{{Dark confinement-deconfinement phase
  transition: a roadmap from Polyakov loop models to gravitational waves}},
  \href{https://doi.org/10.1007/JHEP09(2021)060}{\emph{JHEP} {\bfseries 09}
  (2021) 060} [\href{https://arxiv.org/abs/2101.03795}{{\ttfamily
  2101.03795}}].

\bibitem{Halverson:2016nfq}
J.~Halverson, B.~D. Nelson and F.~Ruehle, \emph{{String Theory and the Dark
  Glueball Problem}},
  \href{https://doi.org/10.1103/PhysRevD.95.043527}{\emph{Phys. Rev. D}
  {\bfseries 95} (2017) 043527}
  [\href{https://arxiv.org/abs/1609.02151}{{\ttfamily 1609.02151}}].

\bibitem{Sannino:2002wb}
F.~Sannino, \emph{{Polyakov loops versus hadronic states}},
  \href{https://doi.org/10.1103/PhysRevD.66.034013}{\emph{Phys. Rev. D}
  {\bfseries 66} (2002) 034013}
  [\href{https://arxiv.org/abs/hep-ph/0204174}{{\ttfamily hep-ph/0204174}}].

\bibitem{Boyd:1996bx}
G.~Boyd, J.~Engels, F.~Karsch, E.~Laermann, C.~Legeland, M.~Lutgemeier and
  B.~Petersson, \emph{{Thermodynamics of SU(3) lattice gauge theory}},
  \href{https://doi.org/10.1016/0550-3213(96)00170-8}{\emph{Nucl. Phys. B}
  {\bfseries 469} (1996) 419}
  [\href{https://arxiv.org/abs/hep-lat/9602007}{{\ttfamily hep-lat/9602007}}].

\bibitem{CP-PACS:1999eop}
{\scshape CP-PACS} Collaboration, M.~Okamoto et~al., \emph{{Equation of state
  for pure SU(3) gauge theory with renormalization group improved action}},
  \href{https://doi.org/10.1103/PhysRevD.60.094510}{\emph{Phys. Rev. D}
  {\bfseries 60} (1999) 094510}
  [\href{https://arxiv.org/abs/hep-lat/9905005}{{\ttfamily hep-lat/9905005}}].

\bibitem{Meisinger:2001cq}
P.~N. Meisinger, T.~R. Miller and M.~C. Ogilvie, \emph{{Phenomenological
  equations of state for the quark gluon plasma}},
  \href{https://doi.org/10.1103/PhysRevD.65.034009}{\emph{Phys. Rev. D}
  {\bfseries 65} (2002) 034009}
  [\href{https://arxiv.org/abs/hep-ph/0108009}{{\ttfamily hep-ph/0108009}}].

\bibitem{Meisinger:2001fi}
P.~N. Meisinger and M.~C. Ogilvie, \emph{{Complete high temperature expansions
  for one loop finite temperature effects}},
  \href{https://doi.org/10.1103/PhysRevD.65.056013}{\emph{Phys. Rev. D}
  {\bfseries 65} (2002) 056013}
  [\href{https://arxiv.org/abs/hep-ph/0108026}{{\ttfamily hep-ph/0108026}}].

\bibitem{Dumitru:2000in}
A.~Dumitru and R.~D. Pisarski, \emph{{Event-by-event fluctuations from decay of
  a Polyakov loop condensate}},
  \href{https://doi.org/10.1016/S0370-2693(01)00286-6}{\emph{Phys. Lett. B}
  {\bfseries 504} (2001) 282}
  [\href{https://arxiv.org/abs/hep-ph/0010083}{{\ttfamily hep-ph/0010083}}].

\bibitem{Agasian:1993fn}
N.~O. Agasian, \emph{{Dilaton at nonzero temperature and deconfinement in
  gluodynamics}}, {\emph{JETP Lett.} {\bfseries 57} (1993) 208}.

\bibitem{Campbell:1990ak}
B.~A. Campbell, J.~R. Ellis and K.~A. Olive, \emph{{{QCD} Phase Transitions in
  an Effective Field Theory}},
  \href{https://doi.org/10.1016/0550-3213(90)90608-G}{\emph{Nucl. Phys. B}
  {\bfseries 345} (1990) 57}.

\bibitem{Simonov:1992bc}
Y.~A. Simonov, \emph{{Calculating deconfinement temperature through the scale
  anomaly in gluodynamics}}, {\emph{JETP Lett.} {\bfseries 55} (1992) 627}.

\bibitem{Sollfrank:1994du}
J.~Sollfrank and U.~W. Heinz, \emph{{Chiral and gluon condensates at finite
  temperature}}, \href{https://doi.org/10.1007/BF01571311}{\emph{Z. Phys. C}
  {\bfseries 65} (1995) 111}
  [\href{https://arxiv.org/abs/nucl-th/9406014}{{\ttfamily nucl-th/9406014}}].

\bibitem{Carter:1998ti}
G.~W. Carter, O.~Scavenius, I.~N. Mishustin and P.~J. Ellis, \emph{{An
  Effective model for hot gluodynamics}},
  \href{https://doi.org/10.1103/PhysRevC.61.045206}{\emph{Phys. Rev. C}
  {\bfseries 61} (2000) 045206}
  [\href{https://arxiv.org/abs/nucl-th/9812014}{{\ttfamily nucl-th/9812014}}].

\bibitem{Schaefer:2001cn}
B.~J. Schaefer, O.~Bohr and J.~Wambach, \emph{{Finite temperature gluon
  condensate with renormalization group flow equations}},
  \href{https://doi.org/10.1103/PhysRevD.65.105008}{\emph{Phys. Rev. D}
  {\bfseries 65} (2002) 105008}
  [\href{https://arxiv.org/abs/hep-th/0112087}{{\ttfamily hep-th/0112087}}].

\bibitem{Drago:2001gd}
A.~Drago, M.~Gibilisco and C.~Ratti, \emph{{Evaporation of the gluon
  condensate: A Model for pure gauge SU(3)(c) phase transition}},
  \href{https://doi.org/10.1016/j.nuclphysa.2004.06.017}{\emph{Nucl. Phys. A}
  {\bfseries 742} (2004) 165}
  [\href{https://arxiv.org/abs/hep-ph/0112282}{{\ttfamily hep-ph/0112282}}].

\bibitem{Renk:2002md}
T.~Renk, R.~Schneider and W.~Weise, \emph{{Phases of QCD, thermal
  quasiparticles and dilepton radiation from a fireball}},
  \href{https://doi.org/10.1103/PhysRevC.66.014902}{\emph{Phys. Rev. C}
  {\bfseries 66} (2002) 014902}
  [\href{https://arxiv.org/abs/hep-ph/0201048}{{\ttfamily hep-ph/0201048}}].

\bibitem{Pisarski:2001pe}
R.~D. Pisarski, \emph{{Tests of the Polyakov loops model}},
  \href{https://doi.org/10.1016/S0375-9474(02)00699-1}{\emph{Nucl. Phys. A}
  {\bfseries 702} (2002) 151}
  [\href{https://arxiv.org/abs/hep-ph/0112037}{{\ttfamily hep-ph/0112037}}].

\bibitem{KorthalsAltes:1999cp}
C.~P. Korthals~Altes, R.~D. Pisarski and A.~Sinkovics, \emph{{The Potential for
  the phase of the Wilson line at nonzero quark density}},
  \href{https://doi.org/10.1103/PhysRevD.61.056007}{\emph{Phys. Rev. D}
  {\bfseries 61} (2000) 056007}
  [\href{https://arxiv.org/abs/hep-ph/9904305}{{\ttfamily hep-ph/9904305}}].

\bibitem{Dumitru:2001xa}
A.~Dumitru and R.~D. Pisarski, \emph{{Degrees of freedom and the deconfining
  phase transition}},
  \href{https://doi.org/10.1016/S0370-2693(01)01424-1}{\emph{Phys. Lett. B}
  {\bfseries 525} (2002) 95}
  [\href{https://arxiv.org/abs/hep-ph/0106176}{{\ttfamily hep-ph/0106176}}].

\bibitem{Wirstam:2001ka}
J.~Wirstam, \emph{{One loop QCD corrections to the thermal Wilson line model}},
  \href{https://doi.org/10.1103/PhysRevD.65.014020}{\emph{Phys. Rev. D}
  {\bfseries 65} (2002) 014020}
  [\href{https://arxiv.org/abs/hep-ph/0106141}{{\ttfamily hep-ph/0106141}}].

\bibitem{Laine:1999hh}
M.~Laine and O.~Philipsen, \emph{{The Nonperturbative QCD Debye mass from a
  Wilson line operator}},
  \href{https://doi.org/10.1016/S0370-2693(99)00641-3}{\emph{Phys. Lett. B}
  {\bfseries 459} (1999) 259}
  [\href{https://arxiv.org/abs/hep-lat/9905004}{{\ttfamily hep-lat/9905004}}].

\bibitem{Sannino:2002re}
F.~Sannino, N.~Marchal and W.~Schafer, \emph{{Partial deconfinement in color
  superconductivity}},
  \href{https://doi.org/10.1103/PhysRevD.66.016007}{\emph{Phys. Rev. D}
  {\bfseries 66} (2002) 016007}
  [\href{https://arxiv.org/abs/hep-ph/0202248}{{\ttfamily hep-ph/0202248}}].

\bibitem{Scavenius:2001pa}
O.~Scavenius, A.~Dumitru and A.~D. Jackson, \emph{{Explosive decomposition in
  ultrarelativistic heavy ion collision}},
  \href{https://doi.org/10.1103/PhysRevLett.87.182302}{\emph{Phys. Rev. Lett.}
  {\bfseries 87} (2001) 182302}
  [\href{https://arxiv.org/abs/hep-ph/0103219}{{\ttfamily hep-ph/0103219}}].

\bibitem{Scavenius:2002ru}
O.~Scavenius, A.~Dumitru and J.~T. Lenaghan, \emph{{The K / pi ratio from
  condensed Polyakov loops}},
  \href{https://doi.org/10.1103/PhysRevC.66.034903}{\emph{Phys. Rev. C}
  {\bfseries 66} (2002) 034903}
  [\href{https://arxiv.org/abs/hep-ph/0201079}{{\ttfamily hep-ph/0201079}}].

\bibitem{DelDebbio:2002nb}
L.~Del~Debbio, A.~Di~Giacomo, B.~Lucini and G.~Paffuti, \emph{{Abelian
  projection in SU(N) gauge theories}},
  \href{https://arxiv.org/abs/hep-lat/0203023}{{\ttfamily hep-lat/0203023}}.

\bibitem{Svetitsky:1982gs}
B.~Svetitsky and L.~G. Yaffe, \emph{{Critical Behavior at Finite Temperature
  Confinement Transitions}},
  \href{https://doi.org/10.1016/0550-3213(82)90172-9}{\emph{Nucl. Phys. B}
  {\bfseries 210} (1982) 423}.

\bibitem{Schechter:1980ak}
J.~Schechter, \emph{{Effective Lagrangian with Two Color Singlet Gluon
  Fields}}, \href{https://doi.org/10.1103/PhysRevD.21.3393}{\emph{Phys. Rev. D}
  {\bfseries 21} (1980) 3393}.

\bibitem{Schechter:2001ts}
J.~Schechter, \emph{{Introduction to effective Lagrangians for QCD}},
  {\emph{eConf} {\bfseries C010815} (2002) 76}
  [\href{https://arxiv.org/abs/hep-ph/0112205}{{\ttfamily hep-ph/0112205}}].

\bibitem{Lucini:2012wq}
B.~Lucini, A.~Rago and E.~Rinaldi, \emph{{SU($N_c$) gauge theories at
  deconfinement}},
  \href{https://doi.org/10.1016/j.physletb.2012.04.070}{\emph{Phys. Lett. B}
  {\bfseries 712} (2012) 279}
  [\href{https://arxiv.org/abs/1202.6684}{{\ttfamily 1202.6684}}].

\bibitem{Gomm:1985ut}
R.~Gomm, P.~Jain, R.~Johnson and J.~Schechter, \emph{{Scale Anomaly and the
  Scalars}}, \href{https://doi.org/10.1103/PhysRevD.33.801}{\emph{Phys. Rev. D}
  {\bfseries 33} (1986) 801}.

\bibitem{Ouyed:2001fv}
R.~Ouyed and F.~Sannino, \emph{{The Glueball sector of two flavor color
  superconductivity}},
  \href{https://doi.org/10.1016/S0370-2693(01)00444-0}{\emph{Phys. Lett. B}
  {\bfseries 511} (2001) 66}
  [\href{https://arxiv.org/abs/hep-ph/0103168}{{\ttfamily hep-ph/0103168}}].

\bibitem{Curtin:2022tou}
D.~Curtin, C.~Gemmell and C.~B. Verhaaren, \emph{{Simulating Glueball
  Production in $N_f = 0$ QCD}},
  \href{https://arxiv.org/abs/2202.12899}{{\ttfamily 2202.12899}}.

\bibitem{Asadi:2021pwo}
P.~Asadi, E.~D. Kramer, E.~Kuflik, G.~W. Ridgway, T.~R. Slatyer and J.~Smirnov,
  \emph{{Thermal squeezeout of dark matter}},
  \href{https://doi.org/10.1103/PhysRevD.104.095013}{\emph{Phys. Rev. D}
  {\bfseries 104} (2021) 095013}
  [\href{https://arxiv.org/abs/2103.09827}{{\ttfamily 2103.09827}}].

\bibitem{Asadi:2022vkc}
P.~Asadi, E.~D. Kramer, E.~Kuflik, T.~R. Slatyer and J.~Smirnov,
  \emph{{Glueballs in a thermal squeezeout model}},
  \href{https://doi.org/10.1007/JHEP07(2022)006}{\emph{JHEP} {\bfseries 07}
  (2022) 006} [\href{https://arxiv.org/abs/2203.15813}{{\ttfamily
  2203.15813}}].

\bibitem{DElia:2002hkf}
M.~D'Elia, A.~Di~Giacomo and E.~Meggiolaro, \emph{{Gauge invariant field
  strength correlators in pure Yang-Mills and full QCD at finite temperature}},
  \href{https://doi.org/10.1103/PhysRevD.67.114504}{\emph{Phys. Rev. D}
  {\bfseries 67} (2003) 114504}
  [\href{https://arxiv.org/abs/hep-lat/0205018}{{\ttfamily hep-lat/0205018}}].

\bibitem{Yamanaka:2019aeq}
N.~Yamanaka, H.~Iida, A.~Nakamura and M.~Wakayama, \emph{{Dark matter
  scattering cross section and dynamics in dark Yang-Mills theory}},
  \href{https://doi.org/10.1016/j.physletb.2020.136056}{\emph{Phys. Lett. B}
  {\bfseries 813} (2021) 136056}
  [\href{https://arxiv.org/abs/1910.01440}{{\ttfamily 1910.01440}}].

\bibitem{Yamanaka:2019yek}
N.~Yamanaka, H.~Iida, A.~Nakamura and M.~Wakayama, \emph{{Glueball scattering
  cross section in lattice SU(2) Yang-Mills theory}},
  \href{https://doi.org/10.1103/PhysRevD.102.054507}{\emph{Phys. Rev. D}
  {\bfseries 102} (2020) 054507}
  [\href{https://arxiv.org/abs/1910.07756}{{\ttfamily 1910.07756}}].

\bibitem{Planck:2018vyg}
{\scshape Planck} Collaboration, N.~Aghanim et~al., \emph{{Planck 2018 results.
  VI. Cosmological parameters}},
  \href{https://doi.org/10.1051/0004-6361/201833910}{\emph{Astron. Astrophys.}
  {\bfseries 641} (2020) A6}
  [\href{https://arxiv.org/abs/1807.06209}{{\ttfamily 1807.06209}}]. [Erratum:
  Astron.Astrophys. 652, C4 (2021)].

\end{thebibliography}\endgroup

\end{document}